\documentclass[a4paper,11pt]{article}

\usepackage{palatino}
\usepackage{subcaption}
\usepackage{amsthm}
\usepackage{pifont}
\usepackage{marvosym}
\usepackage{float}
\usepackage{graphicx}

\title{Computer-Supported Risk Identification for the Holistic Management
       of Risks}

\date{2015-10-27}

\author{Jochen L.~Leidner, Ph.D.\\Director of Research\\Thomson Reuters}

\begin{document}

\maketitle
\tableofcontents
\newpage


\begin{abstract}
  Risk is part of the fabric of every business; surprisingly, there is little
  work on establishing best practices for systematic, repeatable risk
  identification, arguably the first step of any risk management process.
  In this paper, we present a proposal that constitutes a more holistic risk
  management approach, a methodology for computer-supported risk identification
  is proposed that may lead to more consistent (objective, repeatable) risk
  analysis.
\end{abstract}


\section{Introduction}

Pursuing any kind of business activity is inseparably interwoven with being
exposed to different kinds of risk
\cite{Beck:1992,Adams:1995,Bernstein:1998,Taleb:2007,Gigerenzer:2013}:
Is the customer I am dealing with liquid
and honest, i.e.~can I rely on being paid? Are my vendors delivering my
supplies punctually, and to the quality I need? Am I in compliance with all
applicable laws and regulations (commercial law, health \& safety, financial
reporting, tax, human resources etc.)? Are my products and services still
relevant, or is demand shrinking or are markets disrupted by new inventions or
commoditization of technologies? Are my competitors outperforming my product or
undercutting my pricing? Does my business have the right staff in terms of
skills? Am I setting the right priorities? Is the cash flow positive and are
the profit margins acceptable? Am I exposed to currency exchange risk because
many of my customers are in different currency zones? Are my offices in
countries that are politically stable as well as free from natural disasters
so that they can carry out their business activities in an undisturbed way?

The task of finding the comprehensive set of risks faced by an entity---its
\emph{risk register}---is known as \emph{Risk identification}.

\subsection{Motivation}

Risk identification is the first step in any comprehensive risk management
cycle, and to date it has been carried our for many several reasons:
\begin{itemize}
\item The management of a business genuinely wants to learn about the risks
  that the business may suffer from, as part of business planning, project
  management \cite{PMI:2013} or strategic planning activities, or just for
  day-to-day operational use;
\item the business may be obliged to report risks to a regulator, for example
  in the case of U.S. public companies the Form 10-K filing must be annually
  submitted to the Securities and Exchange Commission (SEC), and it includes a
  section (``ITEM 7A. Quantitative and Qualitative Disclosures About Market Risk'')
  on risks \cite{USCongress:1934};
\item before an acquisition or Initial Public Offering (IPO) material risks
  have to be formally disclosed to potential acquiring entities and potential
  investors/shareholders; or
\item a person looking for a job may want to learn about the risks of a
  potential employer before submitting a formal application to it, to
  ascertain the economic viability of the company and its the adherence to his
  or her ethical standards (or the other way round);
\item a bank may carry out a comprehensive risk analysis in order to establish
  whether or not to extend the credit line for a company that is one of their
  clients;
\item an investment manager may hold a portfolio of companies he or she has
  invested in, and may therefore want to ensure that the investment portfolio
  is risk-balanced: the less overlap there is in the kind of risks that the
  portfolio is exposed to, the better.
\end{itemize}

To date, there has been no automated tool support for the risk identification
phase of the risk management process: traditionally, people have drawn up lists
or spreadsheets of business risks from scratch by convening informal meetings,
typical starting out with a blank sheet. The insufficiency of risk
identification has been pointed out before, notably in the context of SEC
filings, where risks are often obtained from competitors' lists via copy \&
paste. This has a number of disadvantages. First, it is unlikely that a list
created from scratch in one session is comprehensive. Second, the approach
of making up the risk register in a meeting without looking at any data means
the risk register will not be complete: very likely, the risks identified thus
will only be the more obvious cases.

In this paper, we propose a technology that has the potential to provide the
basis for a superior approach, a computer-supported risk identification
process. It accomplishes this by supporting humans with automation help in
eliciting evidence for risk exposure from archives and feeds of trusted prose
text, such as news, earnings call transcripts or brokerage documents.

\subsection{Definitions}

A \emph{risk} is a potential future event or situation that has adversarial
implications; it is the possibility of something bad happening in the future.
A \emph{bad event} is when something that once was just a risk---whether it
was recognized before or not---has \emph{materialized}, i.e.~it has
actually happened. According to this terminology, a risk already incorporates
a potential modality, and therefore it makes \emph{no} sense to speak of a
\emph{potential} risk, as that is already implied in the risk term. Events can
\emph{unfold}, i.e.~they can change their spatio-temporal scope, which
may include other, dependent risks materializing in the process.


\section{Risk Identification}

\subsection{Risk Identification as Part of Project Management}

Traditional best practices for project management describe risk identification
as an early step in a sequence of activities including Risk Management Planning,
Risk Identification, Qualitative Risk Analysis, Quantitative Risk Analysis,
Risk Response Planning and Risk Monitoring and Control \cite{PMI:2013}.
However, while the importance of risk identification is acknowledged,
\emph{automated} tool support often is not addressed \cite{Kerzner:2009,PMI:2013}.

\begin{figure}
  \begin{center}
    \includegraphics[width=.8\textwidth]{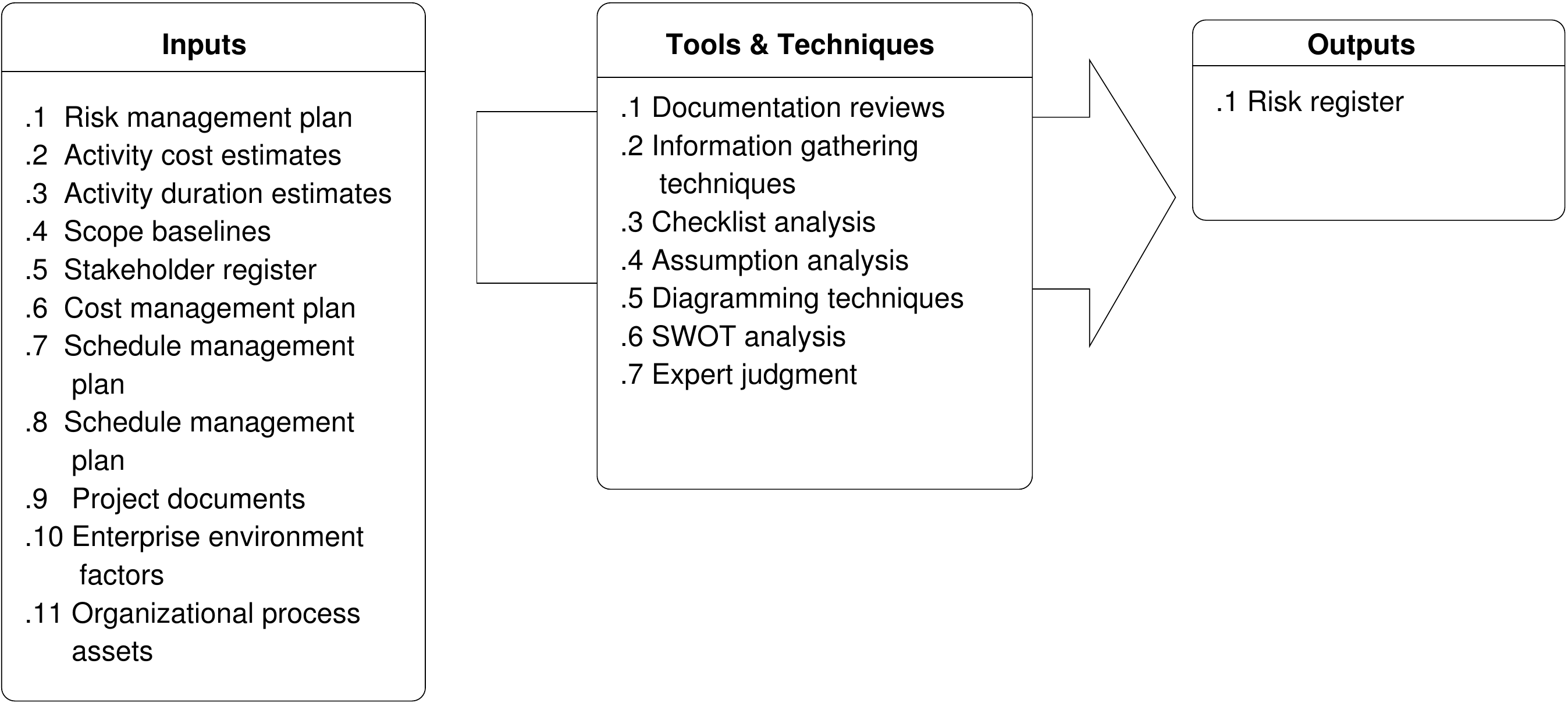}
    \caption{The PMI's \emph{Identify Risks} Process (11--6): Inputs, Tools \& Techniques, and Outputs \cite[Section 11.2]{PMI:2013}.}
    \label{fig:identify-risks}
  \end{center}
\end{figure}

The best practices in project management documented by the Project Management
Institute (PMI) suggest the risk register be generated as output by a set of
tools from a set of inputs~(Fig.~\ref{fig:identify-risks}).
  ``Participants in risk identification activities can include the following:
  project manager, project team members, risk management team (if assigned), customers,
  subject matter experts from outside the project team, end users, other project
  managers, stakeholders, and risk management experts.''~\cite[Section 11.2]{PMI:2013}
Project documents and enterprise environmental factors are listed as inputs, and
they include (ibd., partially cited):
\begin{itemize}
  \item Assumptions log\\[-7mm]
  \item Other project information proven to be valuable in identifying risk\\[-7mm]
  \item Published information, including commercial databases\\[-7mm]
  \item Academic studies\\[-7mm]
  \item Published checklists\\[-7mm]
  \item Industry studies, and\\[-7mm]
  \item Risk attitudes.
\end{itemize}
While some of these sources of evidence may include prose instances of risks,
no mention is made of tool support to \emph{locate} them. In this paper, we
take the position that within large collections of text such as news
archives, risk register elements can lie buried, and that we will need
computer support to unravel them.

Kerzner (2009) recommends ``surveys of the project, customer, and users for
potential concerns''~\cite[p.~755]{Kerzner:2009}, and gives a list of typical
project risks; clearly as of its publication date, automatic tool support for
risk identification was not yet on the horizon, and it is hoped that this
paper will generate initial awareness in favor of automated or semi-automated
methods to collect evidence from textual data.

\subsection{Codification of Risk Identification in Standards}

The International Organization for Standardization (ISO) and the International
Engineering Commission (IEC) have produced a codification of terminology and
best practices for risk management, including risk identification
techniques~\cite{ISO:2009:31000,ISO:2009:31010}. However, because the standard
was issued in 2009, it predates first attempts to develop computerized tools
to support the risk identification stage of the risk management process
\cite{Leidner-Schilder:2010:ACL}.

\subsection{Three Aspects of a Risk}

Key problems of risk management include (1) how to model risk, (2) how to
obtain data for a chosen model so that it can be used in practice, and (3) what
decisions to take based on the risks.

A risk $R$ basically has three properties to characterize it:
\begin{itemize}
\item the \emph{risk type} $R_T$: a name for the description of the risk that
      characterizes the nature of the adversarial potential;
\item a \emph{likelihood} $R_L$: the estimated odds how likely the risk happens
  within a certain time frame (e.g.~6 months) or not;
\item its \emph{impact} $R_I$: \emph{if} it materializes, what is the \emph{severity}
  of the damage caused. This could be expressed as minimum, expected and
  maximum loss in USD, for example, akin to loss databases used by insurances.
\end{itemize}
We can write in short: $R=(R_T,R_L,R_I)$.
Unfortunately, the probability of an event and its impact are often confused by
laypersons and experts alike. A particularly common error is to take the
frequency of mention of a risk type as a proxy for its probability: while in
some cases this makes sense, for example if there are increasingly frequent
reports of political unrest coming from a country, this may indeed be
suggestive of an imminent civil war or revolution, in many cases the
frequent mention of a risk reflects a more extensive, detailed discussion,
which may actually indicate less risk (well scrutinized means better
understood). We will focus on the risk identification step in this paper;
for risk likelihood assessment and impact assessment we refer the reader to the
literature.

\sloppypar{}
Regarding the modeling of risk, one of the easiest approaches is simply listing
the risk types that a company is exposed to, the
\emph{risk register}~(Figure~\ref{fig:risk-register-simple}); at the
most sophisticated end of the spectrum, complex mathematical graph-based models
could simulate propagation of risk evidence, probabilities and causality
through a graph-based model. The simple model does not help you forecast how
likely what will happen, but it permits you to draw up a table of intended
action to deal with each risk type. For the more complex models they may
quantify probability\footnote{
  The ratings ``high'', ``medium'' and ``low'' in Figure~\ref{fig:risk-register-extended} are
  given only for didactic purposes; a real assessment should quantify risk to
  avoid subjective differences in interpreation of these terms \cite[pp.~491-513]{Conrow:2003}.
} and impact, but it is often hard to obtain data for its
parameters, and to validate its appropriateness as a model (are we modeling
the world well?).

\begin{figure}
\centering
\begin{subfigure}{.45\textwidth}
  \centering
    \begin{small}
      \begin{tabular}{|l|} \hline
        \textbf{Acme Inc.} \\ \hline
        \emph{Risk Type} \\ \hline
        office fire risk \\
        cash-flow risk \\
        litigation risk \\
        demand risk \\ \hline
        \end{tabular}
    \end{small}
  \caption{Simple Risk Register.}
  \label{fig:risk-register-simple}
\end{subfigure}%
\begin{subfigure}{.45\textwidth}
  \centering
    \begin{small}
      \begin{tabular}{|l|l|l|l|}\hline
        \multicolumn{4}{|c|}{\textbf{Acme Inc.}} \\ \hline
        \emph{No.} & \emph{Risk Type} & \emph{Likelihood} & \emph{Impact} \\ \hline
        1 & office fire risk          &  low              & high  \\
        2 & cash-flow risk            &  medium           & fatal \\
        3 & litigation risk           &  medium           & high  \\
        4 & demand risk               &  medium           & high  \\ \hline
        \end{tabular}
    \end{small} 
  \caption{Extended Risk Register.}
  \label{fig:risk-register-extended}
\end{subfigure}
\caption{\emph{Risk Register} for a fictional pub\-lishing company. (a) In its
  simplest form, it is a set comprising the list of risk types (left). (b) The
  extended risk register for our fictional company shows the 3 essential
  attributes of each risk $R=(R_T,R_L,R_I)$ (right).}
\label{fig:risk-register}
\end{figure}

\subsection{Evaluating Risk Register Quality}

A risk register's merit can be judged along a couple of dimensions:
\begin{itemize}
\item \emph{comprehensiveness}: does it contain all or at least most risks that
  the entity it pertains to is exposed to? This is difficult because in reality
  we do not have access to the complete universe of risks for an entity to
  compare to.
\item \emph{currency}: does it contain the risks significantly before they
  materialize?
\item \emph{correctness}: how correct are the risks in the risk register? This
  can be measured by Precision, the percentage of correct risks that are also
  present in the risk register. A risk can be deemed correct at different
  levels: at the most basic level, a risk $R_i$ is correctly included in a risk
  register for an entity e if the linguistic context from which the risk
  mention of r was extracted supports the inclusion decision. I.e., more human
  analysts would include, independently from each other, the risk in the risk
  register based on the evidence than those that don't (human agreement is
  always an upper bound of machine performance, at least if machines are
  evaluated against a human ``ground truth'', ``gold standard'' or ``reference
  solution'').
\item \emph{cost}: all things being equal, a risk register is better than
  another if it can be produced more cheaply than another.
\end{itemize}
In the absence of an ``oracle'' that provides the complete set of risks which
could be used for an absolute evaluation, one work-around is to have multiple
systems developed by different groups using different methods for risk
identification, each producing their own risk register for any given entity.
Then the set union of all of them could be formed and reviewed by human
judges, to create a resource that will be defined as the gold standard, and
against which also coverage and recall can be measured to accommodate the
aforementioned ``completeness'' quality criterion. This methodology, known
as \emph{pooling}, has been applied successfully in the evaluation of search
engines at the US National Institute of Standards and Technology~(NIST) in
the Text Retrieval Conferences~(TREC).

\subsection{Of Obvious Risks, Gray Swans and Black Swans}

We can distinguish between three types of risks based on how surprised we would
be if they materialized:\footnote{This is consistent with information theory's
view of surprise as information content (less surpising $\rightarrow$ more
predictible $\rightarrow$ smaller information content).}
\begin{itemize}
\item \emph{Obvious risks}\footnote{
    Taleb also speaks of \emph{White Swans}: ``A White Swan for me would be a
    bridge that can only handle these trucks, of course, and you are certain
    because you have seen from a helicopter a few big six-ton trucks coming on
    the highway, and so you know the bridge is going to collapse, it's only a
    matter of time.'' \cite{Pressler:2010:NYTMag}
} can be important to bring to one's attention (when their
materialization is \emph{imminent}), but often we will want a filter to see
only the not-so-obvious;
\item \emph{Gray Swans} are defined by Taleb \cite{Taleb:2007} as risks that
are hard to anticipate because they are unlikely, and they may have huge impact
once they materialize; and
\item \emph{Black Swan risks} are defined by Taleb \cite{Taleb:2007} as risks
that cannot in principle be anticipated, they have a very low likelihood, yet
their impact is enormous (black swans were believed not to exist until some
very finally discovered). If there exists a class of risks that cannot by
definition be anticipated, it naturally is outside of the scope of
computer-supported techniques for detecting them (which is why we can focus on
``gray swans'' here).
\end{itemize}

\section{``Risk Mining'' from Textual Sources}

\begin{figure}
  \begin{center}
  \includegraphics[width=\textwidth]{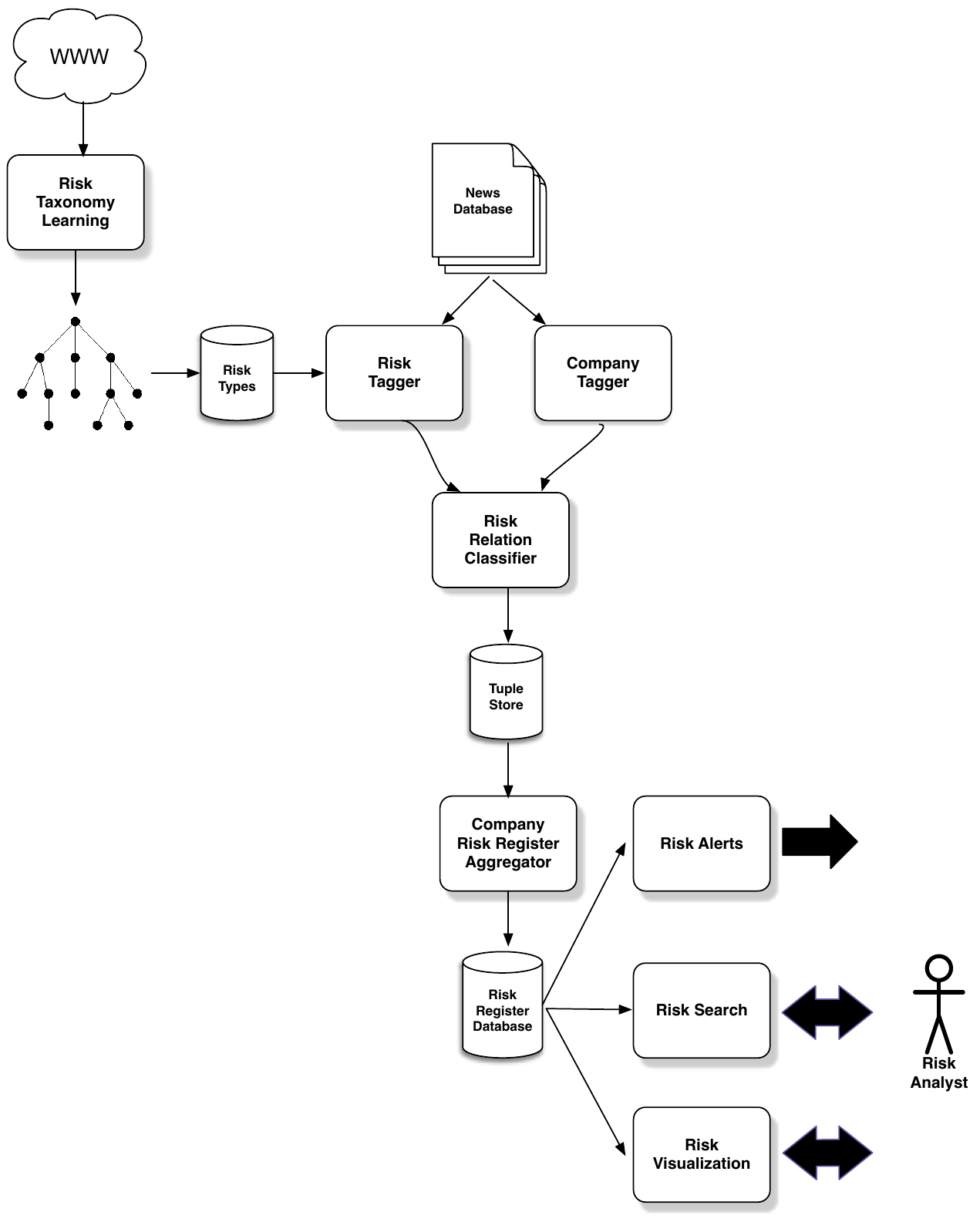}
  \caption{Machine Learning for Computer-Supported Company Risk Exposure Identification.}
  \label{fig:risk-mining-ml}
  \end{center}
\end{figure}

In this section, a method for computer-supported risk identification is
described at the conceptual level (Figure~\ref{fig:risk-mining-ml}); for
more detail, see \cite{Leidner-Schilder:2010:ACL,Nugent-Leidner:2015:TACL}.

The method takes three inputs: (i) the World Wide Web, as indexed by a
search engine, (ii) a set of company names, the risks of which we are
interested in (e.g. a list of suppliers or an investment portfolio) and
(iii) a news archive comprising trusted news and analysis. The World Wide
Web (WWW) is used to mine a taxonomy of risk types, regardless of the
entity that is exposed to them; the WWW was chosen because it is the largest
existing online source of English prose. The news archive is the source of
information, from which we can extract the risks, essentially using
journalists' insights to ``crowdsource'' risk mentions from their
reporting. The company list is the real (variable) input, and the output
is a risk register for each company.

The method comprises of three steps: a taxonomy learning step, which is
run at least once to obtain an inventory of possible names for risks,
a tagging step in which company names and risk type names, respectively,
are annotated in the text of the news feed and/or news archive (by simple
look-up, or possibly by a more sophisticated process such as machine
learning); and a classification step, in which a machine learning process
decides whether a risk mention instance candidate pair comprising a company
name mention and a risk type mention (co-occurring in the same sentence) are
indeed related to each other, and that they indeed express a risk exposure
situation.

\subsection{Machine Learning of Risk Type Taxonomies}

The first step in our method creates a taxonomy of risk terms or phrases,
which we refer to as the \emph{risk taxonomy}. Unlike human-created taxonomy,
the output is very rich in detail, but messy, ``by machines, for machines'' in
a way. We try to obtain a graph with as many IS-A relationships as possible and
``risk'' as its root node by remote-controlling a Web search engine with
search queries for linguistic patterns likely to retrieve risk terms or phrases.

The method, desribed by Leidner and Schilder in more detail
in \cite{Leidner-Schilder:2010:ACL}, makes use of so-called ``Hearst patterns''
(``financial risks such as $\ast$'' is likely to retrieve Web pages, in which
this pattern is followed by ``bankruptcy'', for instance) to induce a rich risk
type vocabulary.

\subsection{Tagging Company Names and Risk Type Names}

Software to automatically annotate prose text with company names is easily
available today. Popular methods are based on name dictionaries (gazetteers),
linguistic rules and/or machine learning.

Likewise, terms and phrases from our risk taxonomy can be tagged or looked up
in sentences.

At the end of this step, each sentence that contains a mention of a company
name and a risk type name has both marked up, which creates candidate pairs
(tuples). Note that the pair \verb|(Microsoft, fine)| could be generated by
both these sentences, one correct and one incorrect (i.e., undesirable in a
risk mining context):

(a) \verb|Microsoft are facing a fine , said Bill Gates .|

(b) \verb|I feel fine , said Microsoft 's Bill Gates .|

What we need is another step that filters out such \emph{false positives}.

\subsection{Machine Learning of Entity-Risk Type Relation Instances}

In order to eliminate spourious false positives in our list of candidate
risk exposure relationship tuples, we can use the method described 
in Nugent and Leidner \cite{Nugent-Leidner:2015:TACL} to
classify each pair comprising a company name and a risk term or phrase,
taking into account the sentential context in which they occur.
For example, supervised machine learning is capable of distinguishing
cases (a) and (b) in the previous section after a few hundred training
sentences have been annotated by human experts to induce a statistical
model from that generalizes the evidence provided in these.

\subsection{Aggregating Risk Registers}

Once risk company-relation mentions have been identified and stored, they can
be \emph{aggregated} so as to form the actual risk register. The naive way of
doing this is by forming the set of all risk mention instance tuples for each
company $C_i$, i.e.~to gather $(C_i, R_j)$ for all $j$s to get the risk register
for one company $C_i$.

\begin{figure}
\centering
\begin{subfigure}{.45\textwidth}
  \centering
    \begin{small}
    \begin{tabular}{|l|} \hline
      \textbf{Acme Inc.} \\ \hline
      \emph{Risk Type} \\ \hline
      office fire risk           \\
      cash-flow risk             \\
      copyright litigation risk   \\
      demand risk                 \\ \hline
      \end{tabular}
    \end{small}
  \caption{ }
  \label{fig:risk-register-qual}
\end{subfigure}%
\begin{subfigure}{.45\textwidth}
  \centering
    \begin{small}
    \begin{tabular}{|lr|}\hline
    \multicolumn{2}{|c|}{\textbf{Acme Inc.}} \\ \hline
    \emph{Risk Type} & \emph{Count} \\ \hline
    office fire risk & 1          \\
    cash-flow risk   & 2         \\
    copyright litigation risk & 1   \\
    demand risk               & 14  \\ \hline
    \end{tabular}
    \caption{ }
    \label{fig:risk-register-quant}
    \end{small}
\end{subfigure}
\caption{\emph{Qualitative} (left) and \emph{quantitative} (right) risk register.}
\label{fig:risk-register}
\end{figure}

Note that a higher frequency indicates merely an increased number of mentions
of a risk, which is not identical, but may in some cases be correlated with,
a higher likelihood for the risk to materialize: a spike in mentions of
``earthquake'' is likely to result from imminent or actual earthquakes, but a
spike in ``acquisition'' may or not preceed the acquisition of a company;
some risks are less likely to materialize just because they are mentioned often,
and that is because all public focus is on the topic, so the risk is at least
not overlooked.

Once a risk register is aggregated, it can be shown to a human analyst for
his or her perusal; however, it makes sense to regularly update the risk
register as part of the Risk Monitoring and Control activity \cite{PMI:2013}
based on new relationships mined that may not have been seen by the system
before.

By retrieving mentions of risks related to companies, risk mining from text
supports the three goals of risk measurement according to
Coleman~\cite{Coleman:2011}: (1) uncovering ``known'' risks,
(2) making the known risks easier to see, and
(3) trying to understand and uncover the ``unknown'' or unanticipated risks.

\subsection{Case Study: Starbucks Corporation}

At the time of writing, Starbucks Corporation is a US-American coffee company
that is operating coffee retail stores internationally.
\emph{Civil unrest risk} is perhaps not the most obvious risk type associated
with this venture, yet our system, Thomson Reuters Risk
Identifier\textsuperscript{TM}, included this risk type in Starbucks' risk
register. Was it an error? Well, the evidence showed that a Starbucks cafe was
used by student protesters as a base to organize their demonstration. If you
think about it, it makes sense, as the perfect place for organizing a demo is
centrally located, has free wireless Internet access, and serves coffee.

Once this risk type is entering the radar of the corporate risk manager of
Starbucks, they can act on it. There are many ways to handle the risk (either
installing house rules that ban demo organizers, or embracing the student
protesters by launching a campaign ``We welcome the student revolution!'');
the point is that it would be unlikely that this kind of risk could be
conceived using traditional risk identification techniques (i.e., a boardroom
meeting with an empty Excel spreadsheet).



\section{Managing Identified Risks}

Once the risk identification, likelihood and impact assessments have concluded,
a risk management plan \cite{PMI:2013} should define the actions to be taken to
influence the risks in the company's favor (Figure~\ref{fig:risk-management-plan}).

\begin{figure}[H]
  \begin{center}
    \begin{small}
      \begin{tabular}{|l|l|}\hline
        \multicolumn{2}{|c|}{\textbf{Acme Inc.}} \\ \hline
        \emph{Risk Type}      & \emph{Management} \\ \hline
        office fire risk      & transfer (buy fire insurance) \\
        cash-flow risk        & mitigate (apply for credit line) \\
        copyright litigation  & mitigate (submit manuscripts to plagiarism checker) \\
        demand risk           & accept (do nothing) \\ \hline
      \end{tabular}
      \caption{A Risk Management Plan.}
      \label{fig:risk-management-plan}
    \end{small}
  \end{center}
\end{figure}


\section{From Individual Risk to Risk Ecosystem}

\subsection{Individual vs Systemic Risk}

Risks can be investigated in isolation; however, quite often, a chain of
follow-up risks is conceivable. Risk-risk connections can be causal or
correlated in nature: if a country is exposed to earthquake risk, then its
citizen may be exposed to hygiene risk since it is likely that water pipes may
burst. The propagation of risk functions regardless of the type of risk,
from hygiene risks to financial risks.

In 1995, Barings Bank failed (caused by unauthorized trading by Nick Leeson,
its head derivatives trader in Singapore) due to particular risks specific
only to Barings (operational risk), whereas the 2008 failure of Lehman
Brothers, AIG, and others was part of a systemic meltdown of global
financial systems caused by bad risk management in the real estate and credit
markets.\footnote{This example is taken from.\cite[Chapter~1]{Coleman:2011}}

Risks can also be inherited from the geo-political environment of operations
when countries are not politically stable~\cite{Henisz-Bennet:2010:HBR} or
ridden by poverty or natural desasters. The World Economic Forum publishes
an annual risk report naming the most pressing global risks.\cite{WEF:2015}

\subsection{Risk Propagation Along The Supply Chain}

\begin{figure}[H]
  \begin{center}
    \includegraphics[width=.8\textwidth]{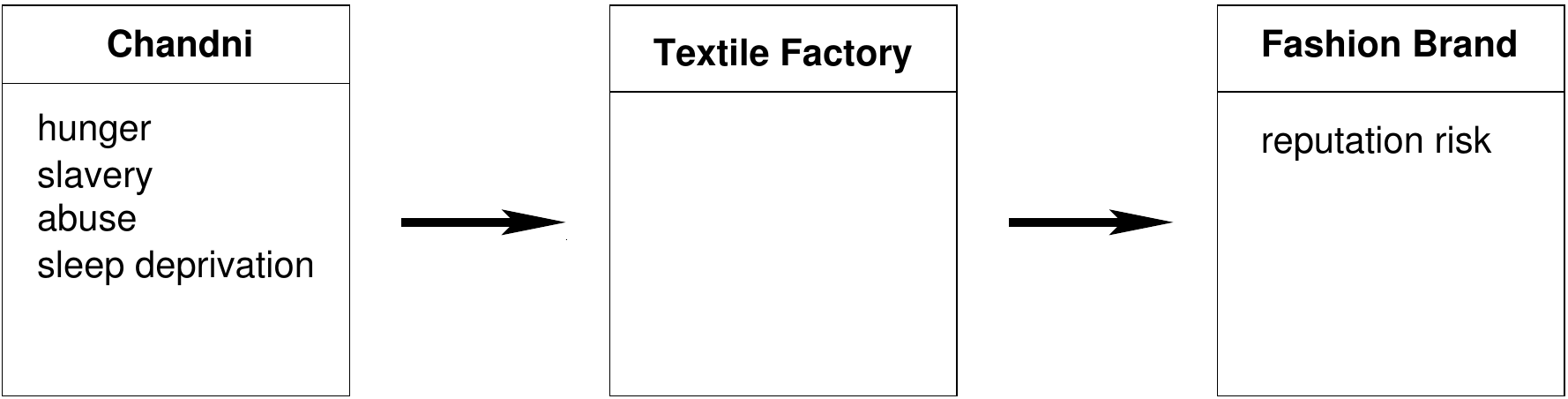}
    \caption{Indirect Reputation Risk Resulting from Worker Explotation in Bangladesh.}
    \label{fig:bangladesh}
  \end{center}
\end{figure}

Imagine Chandni, a textile worker in an old but crowded factory building
(``sweat shop'') in Bangladesh. At the time of writing, she earns \$0.19 per
hour, although she is only twelve years old. She is hungry and lacks sleep, but
kept like a slave, forced to work long hours, and locked in the factory so she
cannot leave.
Figure~\ref{fig:bangladesh} shows how Chandni's personal risk register does not
affect her direct employer much (if she dies, there will be another likely
victim replacing her at the same cost), but the human rights violations she
faces can become a reputation risk for the international fashion brand that
subcontracted the textile factory that employs her.\footnote{While Chandni's
name is made up, the example was inspired by a documentary showing real cases
similar to hers. Location, salary and work conditions are unfortunately not
fictional.}

The suicides of several employees of Foxconn (also known as Hon Hai Precision
Industry Co., Ltd.), an electronics manufacturer that
is a subcontractor for Apple Inc.~(inter alia), has been a prime example of
reputation damage \emph{by association}. Foxconn was reported to exploit its
workers, and some of them took their lives. This in turn caused outrage by
Apple Inc.'s customers when reported by news media \cite{Standing:2010:Reuters}\footnote{
  The truth is often much more complex than journalistic headlines suggest; according
  to Wikipedia (``Foxconn''), the suicide rate of Foxconn is actually below national
  average in China. Note, however, that the fact whether or not   Foxconn's workers'
  suicides is above or below national average does not undo the reputation damage
  Foxconn, and by implication Apple, suffered.
}.

\begin{figure}[H]
  \begin{center}
    \includegraphics[width=.8\textwidth]{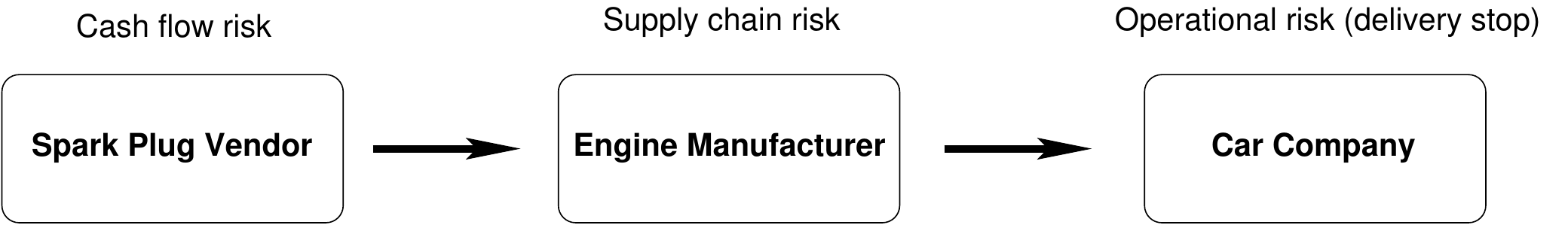}
    \caption{Supply Chain Risk: Risks Must be Propagated along the Supply
             Chain Graph to Account for Transitivity.}
    \label{fig:supply-chain-risk}
  \end{center}
\end{figure}

Another example is a manufacturer of cars, which may source its engines from
a vendor. The engine may contain spark plugs from yet another vendor. If the
spark plug vendor produces a very customized version for the engine manufacturer
that cannot easily be replaced, a cash-flow problem of the spark plug vendor may
delay or even halt production for the car manufacturer if no alternatives can
be sourced easily. The more remote and indirect in the supply chain graph the
risk is from the company that is ultimately (transitively) exposed, the harder
it is to anticipate the problem in the risk identification process. A solution
could be the overlaying of risk registers onto the supply chain
graph~(Figure~\ref{fig:supply-chain-risk}).

\subsection{Risk Model and Real World}

For risk modeling to work very well, it ought to be connected to the real world;
in the risk case, such a ``calibration'' means the model is more grounded in
physical reality, and that it is aligned with actual truth as opposed to mere
predictions, which ultimately are a form of fiction. A risk model that is
informed by real-life signals like loss databases (e.g.~from the insurance
sector), project management databases (as gathered by the project management
offices in corporations) will compare favorable to one that is not linked to
the business operation.

Ideally this connection between risk model and risk reality is bidirectional:
the world informs the model, the model makes predictions, predictions are
compared with real outcomes as risk do or do not materialize, and outcomes are
fed back to improve models. For example, an identified cash flow risk could be
measured legal by how small we permit cash reserves to become, and by comparing
the current balance to the lowest previous low. Or, when identifying legal risk,
we ought to measure actual legal services and litigation cost and feed it back
into our model.

For an organization to be effective, risk modeling and risk management cannot
operate separately from other parts of the business (financial, legal,
operational departments).

\subsection{Portfolio Risk}

Given two publishing companies, Acme Inc.~and Rainforest Publishing Inc., they
will have very different risk registers (Fig.~\ref{fig:portof}; they share the
risks common to all publishing companies, but there will be a set of risks
peculiar to individual companies based on their unique name (e.g.~trademark
violation risk), location (demand risk), pricing (competitive risk), kind of
publications offered (sourcing risk, demand risk), advertising and marketing
mix (operational risk), and so on.

Ed G.~Reedy is an investment manager in charge of 250 million US\$ investment
assets. At any time, he holds a portfolio of securities (e.g.~shares, options,
forward contracts), which make him a stakeholder in the well-being, and
therefore also in the risk exposure, of the underlying companies that make up
his portfolio.

His portfolio comprises five companies, each exposed to a number of partly
different, partly overlapping risks (Fig.~\ref{fig:portfolio-risk}), and it
was assembled in a way that ensures the companies have
high-growth potential, and their risk as far as ``fundamentals''
(financial base numbers like revenue, EBITDA etc.) are not strongly
correlated \cite{Malz:2011}. Once we can look at the risk register for a company
(after extracting it from news text using risk mining, and having
the output vetted by a human analyst), we can scrutinize the
portfolio risk based on the qualitative risk types (as opposed to
scrutinizing it based on fundamentals-based correlation only) by
looking at risk overlap, to get a different perspektive on risk.\footnote{
  \cite{Tetlock-Gardner:2015} suggests that including a diversity of
  sources of evidence (expert \emph{diversity}) is one of the
  cornerstones of the best practices of forecasting.
}

\begin{figure}
\begin{tabular}{|l|r|r|r|r|r|r|r|}\hline
                      & Risk A     & Risk B     & Risk C & Risk D & \,...\, & Risk J & Risk K \\ \hline
  Company A           &            & \ding{122} &        &        &     & \ding{122} &        \\
  Company B           & \ding{122} & \ding{122} &        &        &     &        &        \\
  Company C           &            &            & \ding{122} &        &     & \ding{122}       &        \\
  Company D           &            &            &        & \ding{122} &     &        &        \\
  Company E           & \ding{122} &            &        &        &     &        & \ding{122}       \\   \hline
\end{tabular}
\caption{Portfolio Risk: Sets of Risk Registers Are Better When They Do Not Overlap a Lot.}
\label{fig:portfolio-risk}
\end{figure}

\subsection{Collective Behavior and Regulatory Impact}

\begin{figure}
  \begin{center}
    \includegraphics[width=.65\textwidth]{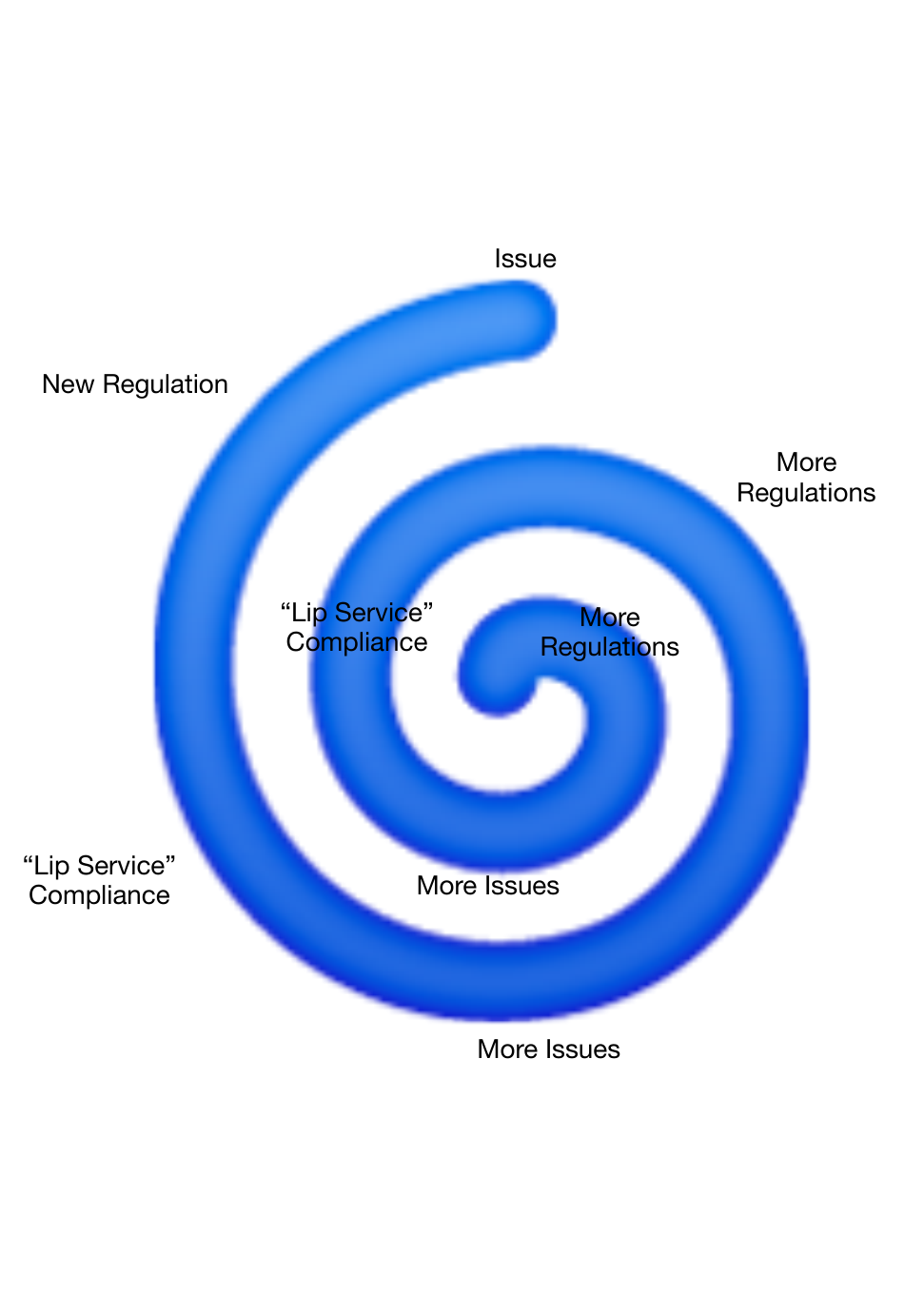}
    \caption{The ``Regulatory Spiral of Death'': Implementing systems for
    the sake of compliance only (i.e., paying only lip service to risk
    management) may lead to systemic breakdown through over-regulation.}
    \label{fig:regulatory-spiral}
  \end{center}
\end{figure}

Companies that pay only lip service to regulation (i.e., they do whatever is
needed to formally comply \emph{in order to comply} rather than in order to
actually reduce the risk exposure of their company) are generating a spiral
of ever-increasing regulatory oversight: because they have only a minimal
interest in actual risk reduction, more new disasters happen. Then based on
markets with dysfunctional self-regulation, regulatory bodies generate more
and stricter regulations and impose them on companies. As a consequence,
the cost of compliance is increased by an attitude of ``mere compliance''.
Of course, subsequent rounds of regulations will cause more expenses in
order to meet even minimal compliance be compliant, which in turn re-inforces
the ``mere compliance'' attitude further. This leads to a spiral of
ever-increasing regulation that may put markets in a regulatory gridlock
resulting in paralysis of the system (Fig.~\ref{fig:regulatory-spiral}).

Therefore, it is crucial to implement risk management for its own sake, and
not just to be compliant. This might require re-aligning, re-vitalizing or
even re-building compliance functions. \cite{Carrel:2010} points out the
importance of culture change in implementing better risk management regimes.


\newpage


\begin{appendix}
\section{Some Notes on Traditional Axiomatic Probability Theory}

The beginning of every kind of probabilistic modeling is the definition of the
set of possible events that make up the total event universe $\Omega$,
consistent with Kolmogorov's axiomatization of
probability~\cite{Kolmogorov:1933}. For example, when dealing with
probabilities of a series of throws of coins $\Omega=\left\{ Head; Tail \right\}$.
However, it is an essential---even \emph{defining}---property of each model to
leave out parts of reality, and what exactly is included or not can vitally
contribute to a model's success. For example, a ``better'' (in the sense of
more complete) model could include the---admittedly extremely rare---case
that a coin lands on neither head nor tail, but on its side:
$\Omega=\left\{ Head; Side; Tail \right\}$.
In the case of the toss of a coin, we may disagree on the utility of including
more and more (very unlikely) cases\footnote{
  The coin could land on its edge, or it could fall from the table.
} in the definition of the event space
$\Omega$, but in risk identification in general, spanning up a very
fine-grained or detailed universe or inventory of possibilities is actually
crucial: if excluded, a hypothetical outcome cannot be even assigned \emph{any}
probability (not even $p=0$) in the later stages.


The traditional definition of $\Omega$ is as a \emph{static} set, and this
implies that either an omniscient vantage point is required for who defines
it, or there will be disastrous consequences of omission. An alternative way to
set up probability theory could conceivably define $\Omega(t)$ as a function of
time, which incrementally grows and also gets more fragmented as new evidence
(about what can happen) surfaces. Such a \emph{dynamic} version of probability
theory, especially under a Bayesian\footnote{based on belief revision} interpretation
of probability, could potentially account for ``knowledge updates'' of $\Omega$.
Imagine this sequence of events:\\
1. $t=0$ (Start): $\Omega(t):=\emptyset$\\
2. $t=1:$ A tossed coin lands on ``Tail'' $\rightarrow$ $\Omega(t):=\left\{ Tail \right\}$ \\
3. $t=2:$ Another toss of the same coin produces ``Head'' $\rightarrow$ $\Omega(t):=\left\{ Tail; Head \right\}$ \\
... (Lots of events, but always either ``Head'' or ``Tail'') \\
1,000,000. $t=999,999: $ Another toss of the same coin produces ``Tail''
$\rightarrow$ $\Omega(t):=\left\{ Tail; Head \right\}$ (i.e., no further updates of the set occurred). \ding{122}
\end{appendix}


\section*{Acknowledgements}
The author gratefully acknowledges discussions on the topic of risk with Frank
Schilder, Khalid Al-Kofahi, Tim Nugent, Chris Brew, Shaun Sibley and Nayeem Sayed.

\newpage


\section*{About the Author}

\begin{small}
Jochen L.~Leidner, Ph.D.~is a Director of Research with Thomson Reuters, where
he heads up the R\&D group in London. He holds a Ph.D.~in Informatics from the
University of Edinburgh, a Master's in Computer Speech, Text and Internet
Technologies from the University of Cambridge and a Master's in Computational
Linguistics, English Lanugage and Literature and Computer Science from
Friedrich-Alexander Universit\"{a}t Erlangen-N\"{u}rnberg.
Prior to Thomson Reuters, he held a Royal Society of Edinburgh Enterprise
Fellowship in Electronic Markets for his work on mobile question answering at
the University of Edinburgh. In the past, he also held roles as a software
developer at SAP AG, and served as a paramedic with the German Red Cross.
He is a certified project management professional (PMP) and member of PMI, IEEE,
ACL and ACM.
\end{small}


\bibliography{risk-mining}

\end{document}